\documentclass[10pt,conference]{IEEEtran}
\IEEEoverridecommandlockouts

\usepackage{lipsum}
\usepackage{enumitem}
\usepackage{xspace}
\usepackage{booktabs}
\usepackage{listings}
\usepackage{courier}
\usepackage{graphicx}
\usepackage{xcolor}
\usepackage{array}
\usepackage{tabularx}
\usepackage{url}
\usepackage{amsmath}
\usepackage{hyperref}
\usepackage{enumitem}
\usepackage{pgfplots}
\usepackage{algorithm}
\usepackage{algpseudocode}
\usepackage{fp}
\usepackage{placeins}

\definecolor{codegreen}{rgb}{0,0.6,0}
\definecolor{codegray}{rgb}{0.5,0.5,0.5}
\definecolor{codepurple}{rgb}{0.58,0,0.82}
\definecolor{backcolour}{rgb}{0.95,0.95,0.92}

\lstdefinestyle{mystyle}{
    backgroundcolor=\color{backcolour},   
    commentstyle=\color{codegreen},
    keywordstyle=\color{magenta},
    numberstyle=\tiny\color{codegray},
    stringstyle=\color{codepurple},
    basicstyle=\ttfamily\footnotesize,
    breakatwhitespace=false,         
    breaklines=true,                 
    captionpos=b,                    
    keepspaces=true,                 
    numbers=left,                    
    numbersep=5pt,                  
    showspaces=false,                
    showstringspaces=false,
    showtabs=false,                  
    tabsize=2
}

\lstdefinestyle{cpp}{
    style=mystyle,
    language=C++,
    keywordstyle=\color{blue},
    keywords=[2]{std},
    keywordstyle=[2]\color{teal},
    keywords=[3]{string, vector, map, set},
    keywordstyle=[3]\color{violet},
    morekeywords={override, final, constexpr, noexcept},
    morecomment=[l][\color{magenta}]{\#}
}

\lstdefinestyle{promptstyle}{
    backgroundcolor=\color{gray!10},
    basicstyle=\ttfamily\footnotesize,
    commentstyle=\color{codegreen},
    keywordstyle=\color{magenta},
    stringstyle=\color{codepurple},
    breakatwhitespace=false,
    breaklines=true,
    captionpos=b,
    keepspaces=true,
    showspaces=false,
    showstringspaces=false,
    showtabs=false,
    tabsize=2,
    frame=single,
    framesep=5pt,
    framerule=0.4pt,
    rulecolor=\color{gray!40},
    xleftmargin=15pt,
    xrightmargin=15pt,
    aboveskip=10pt,
    belowskip=10pt,
    language=,
    columns=flexible,
    keepspaces=true,
    mathescape=true,
    escapechar=§,
    numbers=none  
}

\pgfplotsset{compat=1.17}

\newcommand{\totalwarnings}{135}
\newcommand{\totalfixes}{110}
\newcommand{\acceptedfixes}{102}
\FPeval{\acceptancerate}{round(100 * \acceptedfixes / \totalfixes, 2)}
\newcommand{\optimal}{59}
\newcommand{\ours}{67}
\newcommand{\baseline}{110}
\FPeval{\oursoverbaseline}{round(100 * (\baseline - \ours)/ \baseline, 2)}
\FPeval{\oursdiffoptimal}{round(100 * (\ours - \optimal) / \optimal, 2)}

\def\product{SAP HANA}

\def\BibTeX{{\rm B\kern-.05em{\sc i\kern-.025em b}\kern-.08em
    T\kern-.1667em\lower.7ex\hbox{E}\kern-.125emX}}
\begin{document}

\title{LLM-Based Repair of C++ Implicit Data Loss Compiler Warnings: An Industrial Case Study}

\author{\IEEEauthorblockN{Chansong You}
\IEEEauthorblockA{\textit{SAP Labs Korea}\\
Seoul, South Korea \\
chansong.you@sap.com}
\and
\IEEEauthorblockN{Hyun Deok Choi}
\IEEEauthorblockA{\textit{SAP Labs Korea}\\
Seoul, South Korea \\
hyun.deok.choi@sap.com}
\and
\IEEEauthorblockN{Jingun Hong}
\IEEEauthorblockA{\textit{SAP Labs Korea}\\
Seoul, South Korea \\
jingun.hong@sap.com}
}

\maketitle
\begin{abstract}
This paper presents a method to automatically fix implicit data loss warnings in large C++ projects using Large Language Models (LLMs). Our approach uses the Language Server Protocol (LSP) to gather context, Tree-sitter to extract relevant code, and LLMs to make decisions and generate fixes. The method evaluates the necessity of range checks concerning performance implications and generates appropriate fixes. We tested this method in a large C++ project, resulting in a \acceptancerate\% acceptance rate of the fixes by human developers during the code review. Our LLM-generated fixes reduced the number of warning fix changes that introduced additional instructions due to range checks and exception handling by \oursoverbaseline\% compared to a baseline fix strategy. This result was \oursdiffoptimal\% behind the optimal solutions created by human developers. These findings demonstrate that our LLM-based approach can reduce the manual effort to address compiler warnings while maintaining code quality and performance in a real-world scenario. Our automated approach shows promise for integration into existing development workflows, potentially improving code maintenance practices in complex C++ software projects.
\end{abstract}


\section{Introduction}

Large-scale C++ projects often accumulate numerous compiler warnings over time, presenting significant challenges for development teams. Among these, implicit data loss warnings are particularly prevalent due to C++'s strong typing and complex type system~\cite{stroustrup2013}. These warnings occur when a value is assigned or passed to a variable of a smaller or less precise type, potentially leading to unexpected behavior or data corruption.

Managing these warnings effectively is crucial for maintaining code quality and preventing potential runtime issues. However, addressing them manually can be time-consuming and error-prone, especially in large codebases. As projects grow in size and complexity, the number of warnings can become overwhelming, leading developers to sometimes ignore them~\cite{static_analysis_tool_icse2013}.

Automated Program Repair (APR) emerges as a promising solution to address these challenges. APR aims to automatically fix software bugs without direct human intervention, aligning well with typical software development practices~\cite{apr_tse_2019}. In the context of compiler warnings, an APR system could analyze the warning, understand the code context, and implement a fix that resolves the issue while maintaining the code's intended functionality.

Recent advancements in APR have been significantly driven by the application of LLMs~\cite{apr_llm_2024}. LLM-based APR leverages AI models trained on vast amounts of code and natural language data, demonstrating remarkable ability to understand and generate code. These models offer numerous advantages, including better understanding of code context, learning from diverse bug-fixing patterns, and adaptability to specific warning types or coding standards~\cite{apr_llm_2023}.

In our study, we leverage LLMs to automatically resolve C++ compiler warnings, focusing particularly on implicit data loss warnings. We implement an approach utilizing Language Server Protocol (LSP) features to efficiently provide relevant type information to the LLM, and employ tree-sitter to extract relevant function blocks. And our main contribution is the introduction of a decision-making mechanism to determine the necessity of additional range checks for each fix, crucial for preventing performance overhead in performance-sensitive projects. To ensure the quality and efficiency of generated fixes, we employ the self-consistency method~\cite{selfconsistency_2023}. Our method, tested on \totalwarnings{} implicit data loss warnings, achieved an acceptance rate of \acceptancerate\% among human developers. Additionally, our approach reduced the number of warning fix changes that introduced additional instructions by \oursoverbaseline\% compared to a baseline fix strategy, while the difference between our approach and optimal solutions by human developers was \oursdiffoptimal\%. These results suggest that our method can help address compiler warnings while considering performance implications in large-scale C++ projects.

\section{Background}

\subsection{Compiler Warnings in C++}

Compiler warnings in C++ serve as crucial indicators of potential issues in code that, while not preventing compilation, may lead to unexpected behavior or vulnerabilities during runtime. These warnings are particularly important in C++ due to the language's complex type system and powerful features that can sometimes lead to subtle errors.

Implicit data loss warnings, a common type of compiler warning in C++, occur when a value is assigned or passed to a variable of a smaller or less precise type. For example:

\begin{lstlisting}[style=cpp]
long long bigNumber = 1000000000000LL;
int smallerNumber = bigNumber; // Implicit data loss warning
\end{lstlisting}

Such warnings are prevalent due to C++'s strong typing and implicit conversion rules. They are often triggered in scenarios involving numeric type conversions, function parameter passing, and return value assignments.~\cite{stroustrup2013}

The management of these warnings in large-scale projects presents significant challenges. As codebases grow, the number of warnings can accumulate, potentially leading developers to ignore them \cite{static_analysis_tool_icse2013} - a situation similar to what is known as "alarm fatigue" in other contexts. This phenomenon can result in important warnings being overlooked, potentially compromising code quality and system reliability.

\subsection{Compiler Warnings in Large-Scale C++ Projects}

\begin{figure}
\centering
\includegraphics[width=\linewidth]{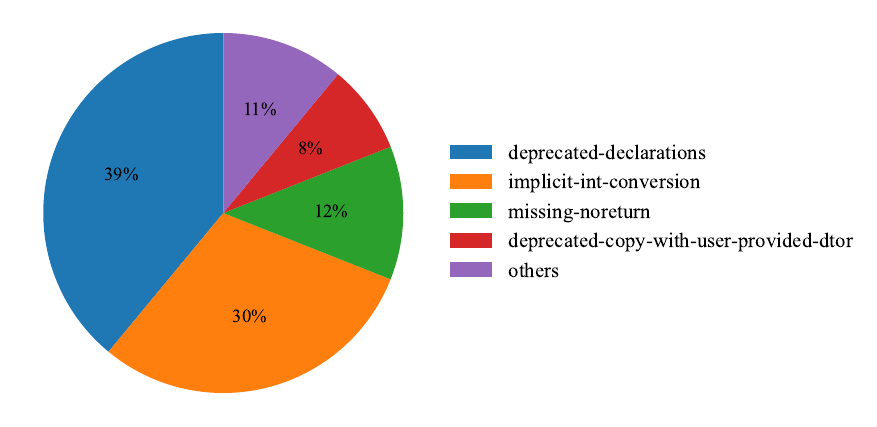}
\caption{Compiler Warning Distribution in SAP HANA}
\label{fig:warnings}
\end{figure}
 
As depicted in Figure \ref{fig:warnings}, SAP HANA's compiler warnings (measured in August 2024) show two main types of issues: deprecated-declarations (39\%) and implicit-int-conversion (30\%), making up about 70\% of all warnings. While deprecated-declarations warnings represent the largest share, addressing them often requires extensive architectural changes and updated interface guidelines, making automated fixes challenging. Our focus is on implicit data loss warnings, a category of compiler warnings that includes implicit-int-conversion warnings.

The remaining warnings are distributed among missing-noreturn (12\%), deprecated-copy-with-user-provided-dtor (8\%), and other miscellaneous warnings (11\%). The high prevalence of implicit-int-conversion warnings (30\%) highlights the significance of addressing implicit data loss warnings in large C++ codebases. These warnings - typically confined to specific variable assignments or function calls - present localized fix patterns that are particularly suitable for automated analysis and repair using LLM-based approaches. While our work focuses on implicit-int-conversion as a concrete example, our solution approach targets the broader category of implicit data loss warnings, providing a comprehensive strategy for handling various types of implicit type conversion issues in C++.

\subsection{Automated Program Repair}

APR is a field in software engineering that aims to automatically fix software bugs without direct human intervention. The primary goal of APR is to reduce the time and effort required for software maintenance, which constitutes a significant portion of software development costs~\cite{apr_tse_2019}.

Traditional APR techniques typically follow a general workflow~\cite{apr_llm_2024}:
\begin{enumerate}
    \item \textbf{Fault localization}: Identifying the likely location of the bug in the code.
    \item \textbf{Patch generation}: Creating potential patches or modifications to address the bug.
    \item \textbf{Patch validation}: Testing the generated fixes to ensure they resolve the issue without introducing new problems.
\end{enumerate}

When applying APR to compiler warnings, particularly in C++, the process is often more straightforward than fixing runtime bugs. Compiler warnings present several advantages for automated repair:
\begin{itemize}[leftmargin=10pt]
    \item The location of the issue is precisely identified by the compiler, eliminating the need for complex fault localization.
    \item Warnings typically include a description of the problem, often suggesting the type of fix required.
    \item The nature of the fix is usually well-defined and localized, often involving simple transformations or additions to the code.
    \item Validation can be done immediately at compile-time, allowing for quick iteration and verification of fixes.
\end{itemize}

For instance, in the case of implicit data loss warnings, fixes might involve explicit type casting, adjusting variable types, or adding range checks. These modifications are generally more predictable and less likely to introduce new bugs compared to fixes for complex runtime issues. Despite the relative simplicity, applying APR to C++ compiler warnings still requires consideration of the language's type system, potential performance implications, and the broader context of the code. However, the defined nature of compiler warnings makes them amenable to automated repair techniques.

\subsection{Large Language Models in APR}

In recent years, the application of LLMs in APR has seen a dramatic rise, marking a significant shift in software engineering practices. The trend began to gain momentum in 2020, with a substantial increase in publications leveraging LLMs for bug-fixing tasks. By 2023, the field saw a surge to 65 papers, indicating growing interest and rapid advancements in this area. Projections suggest this trend will continue, with estimates of over 90 relevant publications expected by the end of 2024. The majority of these studies focus on developing novel repair techniques or methodologies, while others conduct empirical analyses, create benchmarks, or explore practitioner perspectives through surveys. This rapid growth shows the potential of LLMs to improve APR, offering new approaches to long-standing challenges in software maintenance and quality assurance.~\cite{apr_llm_2024}

A notable application of LLMs in APR has been in addressing static warnings, particularly those generated by code analysis tools. Pioneering work in this area includes TFix~\cite{tfix_2021}, which reformulated program repair as a text-to-text task using T5-based models. Subsequent approaches like InferFix~\cite{inferfix_2023} and RAP-Gen~\cite{rapgen_2023} have built upon this foundation, incorporating retrieval-augmented generation techniques to enhance repair capabilities. More recent developments, such as CORE~\cite{core_2023} and FDSP~\cite{fdsp_2024}, have leveraged advanced models like ChatGPT and GPT-4 to not only generate fixes but also to rank and refine them, particularly focusing on code quality issues and security vulnerabilities. These advancements demonstrate the growing sophistication of LLM-based approaches in handling a wide range of static code issues, from general coding errors to specific security concerns, marking a significant step forward in automated code improvement and maintenance.

\begin{figure*}[ht]
\centering
\includegraphics[width=\textwidth]{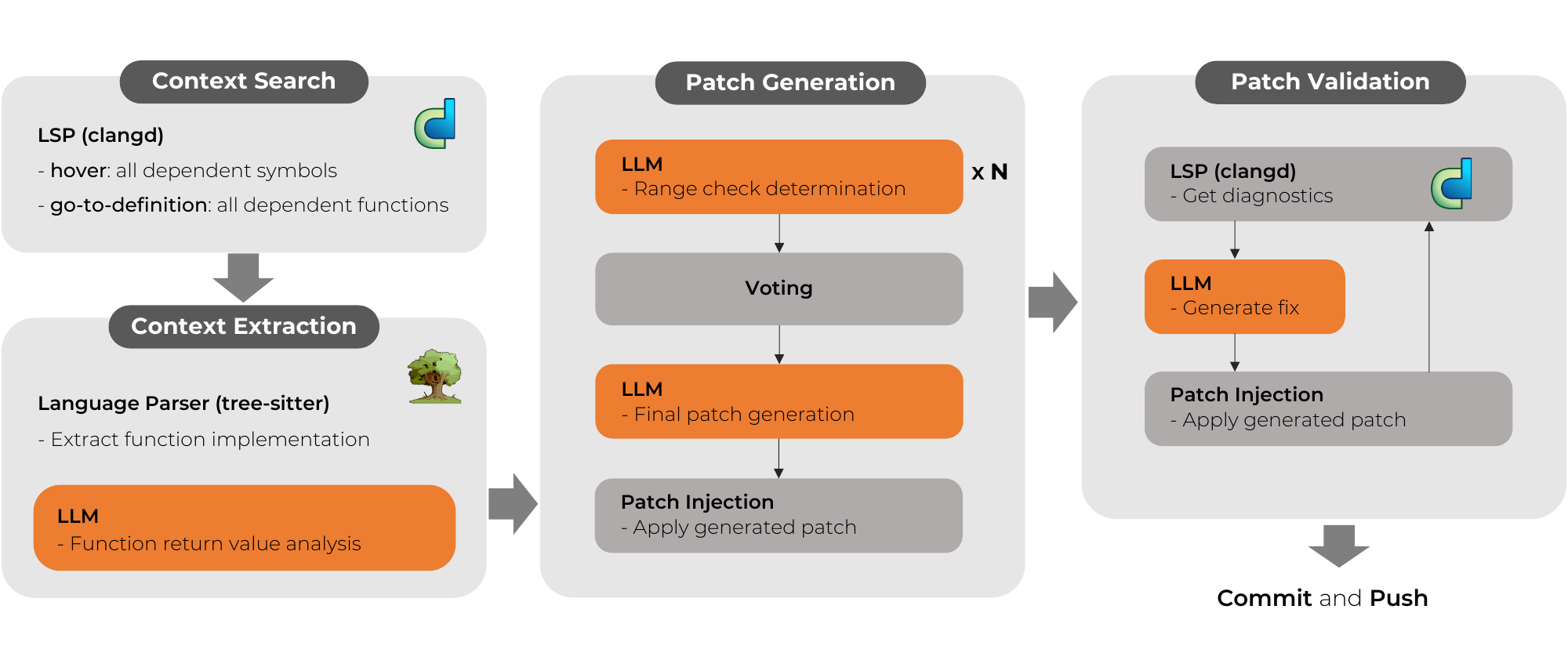}
\caption{System Architecture for Automated Warning Fix Generation}
\label{fig:system_architecture}
\end{figure*}

\section{Methodology}

As described in Figure \ref{fig:system_architecture}, our approach to addressing implicit data loss warnings in large-scale C++ projects involves several key components: context search, context extraction, patch generation and validation. This section details each of these components and explains how they work together to create an effective LLM-driven repair system.

\subsection{Context Search}

The first step in our methodology is gathering relevant context to inform the LLM for generating accurate and effective warning fix patches. To achieve this, we leverage the capabilities of the LSP~\cite{lsp}, a standardized communication protocol between development tools and language-specific analysis servers. Specifically, we use clangd~\cite{clangd}, an LSP implementation for C/C++, as our LSP server.

Our context search methodology primarily utilizes two key LSP features:
\begin{itemize}[leftmargin=10pt]
    \item \textbf{Go-to-definition}: This request allows us to resolve the definition location of symbols at specific text document positions. It provides access to type declarations and function definitions associated with the identifiers involved in the warning.
    \item \textbf{Hover}: This feature supplies concise summaries of types and identifiers, offering additional context without the need to navigate to full definitions.
\end{itemize}

We apply these LSP features to all identifiers present in the line where a compiler warning occurs. Furthermore, if the variable triggering the warning has been used in preceding lines, we extend our context search to include related identifiers' definition locations.

\subsection{Context Extraction}

After gathering relevant context using the LSP, we employ Tree-sitter~\cite{treesitter} for precise code extraction. Using the location information obtained from clangd, we extract the specific function or method containing the warning. Tree-sitter generates a syntax tree for this extracted code segment, allowing us to efficiently navigate and extract relevant parts of the code.

To address the potential cost of including entire function body implementations within the LLM's limited context, we introduce an additional step. Instead of providing the full function body as context, we query the LLM to analyze and explain the function's return value range. We prompt the LLM with the function signature and implementation, asking it to describe the possible range of return values. This approach provides crucial information about function behavior while reducing the amount of context needed. The return value range information is then included in the context provided to the LLM for generating fixes, helping to determine the necessity of range checks and assess the potential for data loss in type conversions.

\subsection{Patch Generation}

Our patch generation process leverages the self-consistency method introduced by Wang et al.~\cite{selfconsistency_2023} to make robust decisions about implementing range checks and generating appropriate fixes.

\subsubsection{Range Check Determination}

To determine whether an additional range check is necessary, we employ the self-consistency approach. This step is important, especially in performance-sensitive parts of the program where constant overhead might be problematic. The process works as follows:

\begin{enumerate}
    \item We prompt the language model with chain-of-thought prompting, asking whether a range check is necessary for the given code conversion.
    
    \item Instead of greedily decoding a single optimal reasoning path, we sample multiple (e.g., 11) diverse reasoning paths from the language model's decoder. Each reasoning path might lead to a different conclusion about the necessity of a range check.
    
    \item We then marginalize out the sampled reasoning paths to find the most consistent answer. Specifically, we take a majority vote over the final decisions (implement range check or not) from all sampled paths.
\end{enumerate}

Here's an example of the prompt structure used for each sampled reasoning path:

\begin{lstlisting}[style=promptstyle]
[System]
You are an expert C++ developer working on a highly performance-sensitive project. Avoid unnecessary operations, even if they seem harmless.

[Code]
long long bigNumber = getSomeValue();
int smallerNumber = bigNumber; // Implicit data loss warning

[Type Information]
bigNumber: long long (64-bit signed integer)
smallerNumber: int (32-bit signed integer)

[Function Signature and Analysis]
int getSomeValue()
Analysis: The function always returns a value between 0 and 9 due to the modulo operation (% 10) applied to the return value.

[Human]
Do we need a range check to resolve this compiler warning?

Think step-by-step:
1. Analyze types and value ranges
2. Assess correctness concerns
3. Evaluate performance implications
4. Come up with an answer of the question with reasoning behind

[Assistant]
1. Types and ranges:
   - bigNumber (long long): 0-9 (from getSomeValue)
   - smallerNumber (int): can represent 0-9

2. Correctness:
   - No data loss risk (0-9 fits in int)

3. Performance:
   - Range check would add unnecessary overhead

4. Recommendation:
   No range check needed. Reasoning:
   - getSomeValue guarantees 0-9 range
   - int can represent all possible values
   - Check would add overhead without benefits

\end{lstlisting}

This self-consistency approach allows us to leverage multiple reasoning paths, reducing the variance of decision-making results. It helps mitigate the impact of any single flawed reasoning path and focuses on the consistency among multiple generated solutions.

\subsubsection{Final Patch Generation}
Based on the range check decision from the self-consistency approach, we generate the final patch. In our example, the decision was that a range check is unnecessary due to the guaranteed small range of values returned by the \lstinline|getSomeValue()| function. In such cases, we guide the LLM to choose between two options:

\begin{enumerate}
    \item Use \lstinline|static_cast| for explicit type conversion.
    \item Modify the type itself, such as using the \lstinline|auto| keyword when applicable, allowing for type inference which can prevent implicit conversions.
\end{enumerate}

The LLM selects the most appropriate option based on the specific context of the warning and the surrounding code. Then, we instruct the LLM to generate its output in a git-diff-like format, where removed lines start with \lstinline|-| and added lines start with \lstinline|+|. For example:

\begin{lstlisting}[style=cpp]
- long long bigNumber = getSomeValue();
+ auto bigNumber = getSomeValue();
  int smallerNumber = bigNumber;
\end{lstlisting}

We introduce a git-diff-like format for the LLM's output to focus on generating only the necessary changes, rather than the entire source code. This approach is particularly beneficial for long function blocks and helps address the limitation of output token limits in current LLM systems. And this output gets processed to apply the changes to the source file where the warning was located, allowing us to resolve compiler warnings by modifying only the relevant parts of the code.

In this example, the LLM has chosen to use \lstinline|auto| for type inference of \lstinline|bigNumber|, as the exact type returned by \lstinline|getSomeValue()| is known to be \lstinline|int| with values in the range 0-9. This change allows the compiler to deduce the correct type (\lstinline|int|) based on the return value of \lstinline|getSomeValue()|. As a result, there is no remaining implicit conversion when assigning to \lstinline|smallerNumber|, as both variables are now of type \lstinline|int|. This approach effectively resolves the initial warning without introducing any unnecessary runtime checks or explicit casts. It maintains both correctness and performance in this performance-sensitive context while simplifying the code by using type inference.

It's worth noting that in cases where a range check is deemed necessary (which wasn't the case in our example), we would instruct the LLM to use a custom type casting utility function that can be implemented as below. This function would incorporate range check and exception throwing, providing a general solution to handle potential data loss scenarios in more complex cases.

\begin{lstlisting}[style=cpp]
template<typename To, typename From>
To safe_int_cast(From value) {
    if (value > std::numeric_limits<To>::max() || 
        value < std::numeric_limits<To>::min()) {
        throw std::runtime_error();
    }
    return static_cast<To>(value);
}
\end{lstlisting}

\subsection{Patch Validation}

\begin{algorithm}[h!]
\caption{Iterative Patch Validation and Correction}
\label{alg:patch_validation}
\begin{algorithmic}[1]
\Procedure{ValidateAndCorrectPatch}{$originalWarning$, $maxIterations$}
\State $patch \gets$ GenerateInitialPatch($originalWarning$)
\State ApplyPatch($patch$)
\State $iterations \gets 0$
\While{$iterations < maxIterations$}
\State $diagnostics \gets$ RunClangdDiagnostics()
\If{$diagnostics$ is empty}
\State \textbf{return} SUCCESS
\Else
\State $newFix \gets$ GenerateFix($diagnostics$)
\State ApplyPatch($newFix$)
\State $iterations \gets iterations + 1$
\EndIf
\EndWhile
\State \textbf{return} FAILURE
\EndProcedure
\end{algorithmic}
\end{algorithm}

After generating and applying a patch, we employ a validation step to ensure the effectiveness of the fix and to catch any new issues that might have been introduced. This process leverages the LSP, specifically clangd's run-time diagnostics capabilities.

Once a patch is applied, we use clangd to analyze the modified code. This allows us to verify whether the original implicit data loss warning has been resolved. However, the patch may sometimes introduce new warnings or even compilation errors. For instance, an LLM-generated \lstinline|static_cast| to a custom type might raise a compilation error due to missing namespace qualifications.

In cases where new issues are detected, we use the detailed diagnostic information provided by clangd. This information is fed back to the LLM, initiating an iterative process, as described in Algorithm~\ref{alg:patch_validation}. It generates and applies an initial patch, then enters a loop that continues until either the maximum number of iterations is reached or a successful fix is found. In each iteration, clangd diagnostics are run on the modified code. If no diagnostics are reported and the original warning is no longer present, the process returns success. Otherwise, a new fix is generated based on the current diagnostics, applied, and the iteration counter is incremented. The final patch not only resolves the original warning, but also maintains the overall integrity of the code.

\section{Experimental Results}

To evaluate the effectiveness of our LLM-driven approach in fixing implicit data loss warnings, we conducted experiments on a large-scale software project, \product. This project is substantial in size, comprising approximately 36M lines of code (LoC)~\cite{bach_2022}. For our study, we focused on a performance-sensitive component of the software that contained \totalwarnings{} implicit-int-conversion warnings. And, to resolve all these warnings, \totalfixes{} patches were required as one fix patch may remove multiple warnings in several cases. 

\subsection{Range Check Decision Making}

\begin{figure*}
\centering
\includegraphics[width=\textwidth]{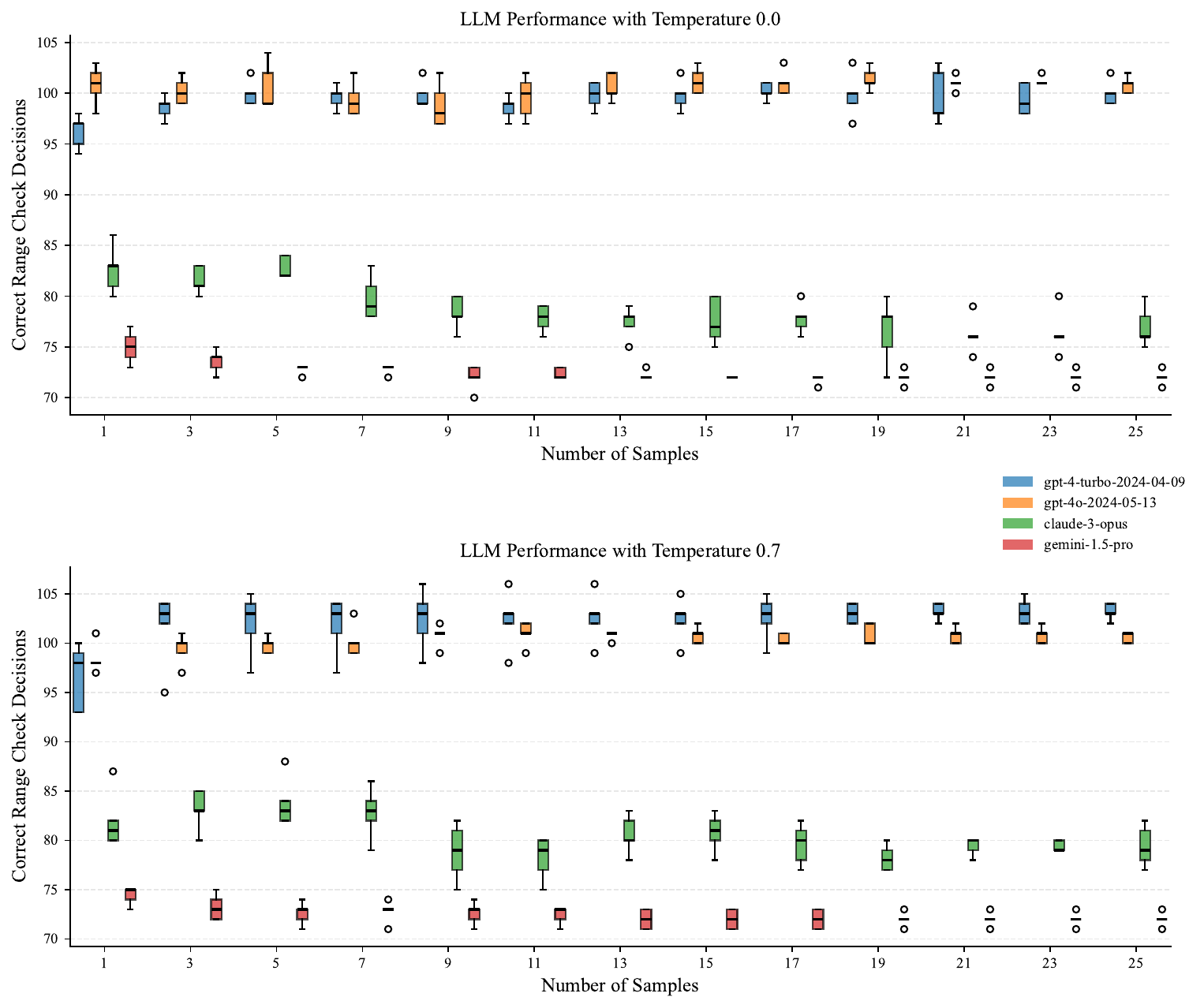}
\caption{Number of correct range check decisions with various LLMs and number of samples used for self-consistency}
\label{fig:llm_correct_responses}
\end{figure*}



Before evaluating the final generated patches, we first focused on measuring the accuracy of the initial range check decision-making, as this fundamental determination directly governs the correctness of the entire fix. This methodological choice was motivated by our observation that, given a correct decision about range check necessity, the subsequent patch generation showed consistent correctness due to the relatively straightforward nature of implementing these fixes in C++.

The experiment utilized four state-of-the-art LLMs: GPT-4 Turbo (2024-04-09), GPT-4o (2024-05-13), Claude-3 Opus, and Gemini-1.5 Pro. For each model, we tested two temperature settings (0.0 and 0.7) to evaluate the impact of sampling randomness on decision quality.
We provided \totalfixes{} code snippets with the total \totalwarnings{} implicit data loss warnings, and each code snippet was presented to the LLMs as a self-contained unit requiring analysis and a decision about range check necessity.
For evaluation, we tested the self-consistency sampling technique, varying the number of samples from 1 to 25 in incremental steps of 2. For each sample size configuration, we repeated the experiment five times to account for potential variability in model responses.

The experimental results demonstrate notable variations in performance across different LLM models and temperature settings for identifying necessary range checks in C++ implicit data loss warnings. Figure~\ref{fig:llm_correct_responses} illustrates the performance patterns across varying numbers of samples (1-25) for both temperature settings (0.0 and 0.7).
GPT-4 based models (GPT-4 Turbo and GPT-4o) consistently outperformed other models, achieving a correct decision count of 95 across different sample sizes. Particularly, GPT-4 Turbo with temperature 0.7 reached optimal performance at relatively small sample sizes, achieving maximum average count of 102 with just 9 samples. This early convergence suggests efficient decision-making capability with minimal self-consistency sampling requirements.

The higher temperature setting (0.7) generally yielded superior results compared to temperature 0.0 across most models. This performance difference was most pronounced in the GPT-4 Turbo model, where temperature 0.7 showed both higher median scores and more stable performance across different sample sizes. This observation suggests that increased sampling diversity in the reasoning paths, facilitated by higher temperature settings, may contribute to more robust decision-making in identifying necessary range checks.

In contrast, Claude-3 Opus and Gemini-1.5 Pro showed relatively consistent but lower performance levels, with correct decision counts ranging between 70 and 85. These models demonstrated less sensitivity to temperature variations, maintaining similar performance patterns across both settings.

The results also indicate that increasing the number of samples beyond 9-11 provided diminishing returns for most models, particularly for the GPT-4 variants. This finding suggests an optimal trade-off point between computational resource utilization and accuracy improvement in self-consistency sampling for this specific task.

\subsection{Fix Acceptance Rate}

\begin{figure}[t]
\centering
\includegraphics[width=\linewidth]{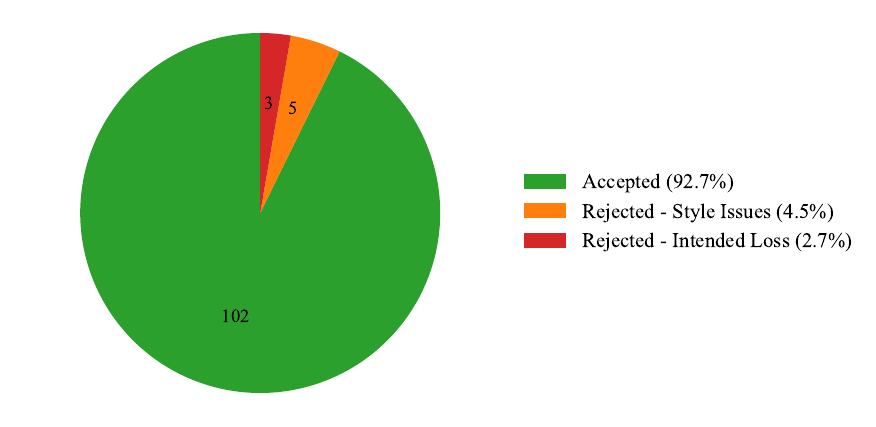}
\caption{Acceptance and rejection rate of LLM generated warning fixes}
\label{fig:fix_acceptance_rate}
\end{figure}

Based on our previous analysis of range check decision accuracy, we selected GPT-4 Turbo with temperature 0.7 and 13 samples for self-consistency as our optimal configuration for generating warning fixes. This configuration achieved the most stable and accurate range check decisions while maintaining reasonable computational efficiency.
As shown in Figure~\ref{fig:fix_acceptance_rate}, using this configuration, we generated fixes for 110 code snippets containing 135 implicit-int-conversion warnings. Our approach achieved a \acceptancerate{}\% acceptance rate, with 102 fixes being accepted through a rigorous code review process. The review was conducted by experienced developers familiar with the project's codebase and performance requirements, who evaluated each fix for correctness, style compliance, and performance implications.

The rejected fixes fell into two distinct categories. Five fixes (4.5\%) were rejected because human reviewers preferred more architectural-level solutions, such as modifying function argument types or broader refactoring, rather than the local range-check fixes proposed by the LLM. While the LLM's fixes were technically correct, reviewers identified opportunities for more comprehensive improvements at the design level. The remaining three fixes (2.7\%) were rejected because they attempted to add range checks for cases where data loss was intentional in the original code design.

The high acceptance rate of \acceptancerate{}\% demonstrates that our approach, particularly the focus on accurate range check decisions before fix generation, effectively produces fixes that not only resolve the warnings but also maintain the project's code quality standards. The analysis of rejected fixes reveals an interesting insight: while LLMs can effectively generate correct local fixes, human developers might sometimes prefer more comprehensive refactoring opportunities that require broader system knowledge and design context.

\subsection{Performance Impact}

Given the performance-sensitive nature of the selected component, we conducted an analysis of the performance impact of our fixes. We used a baseline fix strategy, which involves a type cast utility function containing both a range check and exception-raising mechanism, as our point of comparison. Our goal was to measure the number of warning fix patches that introduced additional instructions by our LLM-generated fixes compared to this baseline approach, as well as to the optimal solution implemented by human developers.

The baseline approach introduces the highest number of fix patches that contain additional instructions (\baseline) due to the comprehensive type cast utility function with range checks and exception handling. Our LLM-generated fixes vary, with some scenarios adding zero instructions by using \lstinline|static_cast| or modifying types directly when no range check is needed, resulting in \ours{} changes with additional instructions. The optimal solution, created by human developers, often includes interface-level modifications that are difficult for the LLM to predict, thus resulting in the lowest number of changes with overhead (\optimal).

\section{Discussion and Conclusion}
Our LLM-driven approach for fixing implicit data loss warnings in large-scale C++ projects demonstrated promising results, achieving a \acceptancerate\% acceptance rate among human developers and reducing warning fix changes with additional instructions by \oursoverbaseline\% compared to the baseline strategy. This performance was close to optimal solutions developed by human experts, highlighting our method's potential to maintain high-performance standards while effectively addressing warnings in critical code sections.

However, our analysis also revealed several limitations of our approach, highlighting areas for future improvement:
\begin{itemize}
    \item \textbf{Design-level changes}: Our system faced challenges in situations where the optimal fix involved modifying function signatures or other high-level design elements. Such changes require a comprehensive understanding of the entire codebase and potential cascading effects, which current LLMs struggle to achieve.
    \item \textbf{C++ macro-generated code}: The LSP doesn't provide accurate context for macro-generated code, limiting our system's effectiveness in these scenarios. This led us to exclude such cases from our experiment, indicating a need for improved handling of macro-expanded code in future iterations.
    \item \textbf{Intended data loss}: In some instances, developers intentionally allow implicit type casting that results in data loss for specific purposes (e.g., generating ID values). Our current approach lacks the capability to identify and preserve these intentional data loss scenarios, highlighting the need for more nuanced understanding of code intent.
\end{itemize}

The results suggest that LLM-driven solutions can contribute significantly to addressing compiler warnings in large-scale C++ projects, balancing automation with code quality and performance maintenance. While our approach shows promise in automating warning resolution, it also highlights the ongoing role of human expertise in software maintenance, especially for complex scenarios requiring deep contextual understanding. Future work should focus on refining these techniques to better handle complex code contexts and integrate software design principles, potentially leading to more efficient software maintenance practices in large C++ projects.

\bibliographystyle{IEEEtran}
\bibliography{references}

\end{document}